\documentstyle{mn}

\title[Dynamics of Neptune Trojan: Eccentric Orbits]
{Dynamics of Neptune's Trojans: II. Eccentric orbits and observed
ones}

\author[Zhou, Dvorak \& Sun]
{Li-Yong Zhou$^{1}$\thanks{zhouly@nju.edu.cn}, Rudolf Dvorak$^2$, Yi-Sui Sun$^{1}$ \\
$^1$Department of Astronomy \& Key Laboratory of Modern Astronomy and Astrophysics in Ministry of Education, \\
  Nanjing University, Nanjing 210093, China\\
$^2$Institute for Astronomy, University of Vienna,
        T\"{u}rkenschanzstr. 17, A-1180 Wien, Austria}

\date{Accepted . Received }
\pagerange{\pageref{firstpage}--\pageref{lastpage}} \pubyear{2009}

\begin{document}

\label{firstpage}

\maketitle

\begin{abstract}
In a previous paper, we have presented a global view of the
stability of Neptune Trojan (NT hereafter) on inclined orbit. As
the continuation of the investigation, we discuss in this paper
the dependence of stability of NT orbits on the eccentricity. For
this task, high-resolution dynamical maps are constructed using
the results of extensive numerical integrations of orbits
initialized on the fine grids of initial semimajor axis ($a_0$)
versus eccentricity ($e_0$). The extensions of regions of stable
orbits on the ($a_0, e_0$)-\,plane at different inclinations are
shown. The maximum eccentricities of stable orbits in three most
stable regions at low ($0^\circ, 12^\circ$), medium ($22^\circ,
36^\circ$) and high ($51^\circ, 59^\circ$) inclination, are found
to be 0.10, 0.12 and 0.04, respectively. The fine structures in
the dynamical maps are described. Via the frequency analysis
method, the mechanisms that portray the dynamical maps are
revealed. The secondary resonances, concerning the frequency of
the librating resonant angle $\lambda-\lambda_8$ and the frequency
of the quasi 2:1 mean motion resonance (MMR for short hereafter)
between Neptune and Uranus, are found deeply involved in the
motion of NTs. Secular resonances are detected and they also
contribute significantly to the triggering of chaos in the motion.
Particularly, the effects of the secular resonance $\nu_8,
\nu_{18}$ are clarified.

We also investigate the orbital stabilities of six observed NTs by
checking the orbits of hundreds clones of them generated within
the observing error bars. We conclude that four of them, except
2001 QR322 and 2005 TO74, are deeply inside the stable region. The
2001 QR322 is in the close vicinity of the most significant
secondary resonance. The 2005 TO74 locates close to the boundary
separating stable orbits from unstable ones, and it may be
influenced by a secular resonance.
\end{abstract}


\section{Introduction}
Besides the trans-Neptunian region (also known as ``Kuiper belt'')
outside the orbit of Neptune and the ``main belt'' in between the
orbits of Mars and Jupiter, the trojan-cloud around some planets
is another main reservoir of minor objects in the Solar system.
Thousands of Trojan asteroids orbiting around the equilateral
Lagrange equilibrium points of Jupiter have been observed and
catalogued since the discovery of (588) Achilles in 1906
\cite{nic61}. In recent years, six NTs were discovered\footnote
{IAU: Minor Planet Center, http://www.cfa.harvard.edu/iau/
lists/NeptuneTrojans.html} and many more of them are expected in
future \cite{she06}. As for other planets, except for Mars, of
which 4 Trojans are known, investigations show the evidences of
large unstable areas around the triangular Lagrange points
\cite{mar03a,nes02a,dvo10}.

Since the observed NTs show a wide range of orbital elements
(especially the inclination, see Table 1) and the potential Trojan
swarm could be much more thick than the one of Jupiter
\cite{she06}, many studies have focused on this asteroid
population, with interests both in the orbital dynamics and
capturing history e.g.,
\cite{bra04,zho09a,zho09b,lij07,nes09,lyk09}.

In a previous paper (Zhou, Dvorak \& Sun 2009, hereafter Paper I),
we investigated the orbital stability of massless artificial
asteroids around the triangular Lagrange points of Neptune ($L_4$
and $L_5$) in a model consisting of the Sun and four outer planets
(Jupiter, Saturn, Uranus and Neptune). We constructed the
``dynamical maps'' on the plane of initial semimajor axis versus
inclination $(a_0, i_0)$ to study the dependence of stability on
inclination. We confirmed two stable regions with inclination in
the ranges of $(0^\circ,12^\circ)$ and $(22^\circ,36^\circ)$
\cite{nes02a}, and found a new stable region for highly-inclined
orbits with $i_0\in(51^\circ, 59^\circ)$.

To complete the investigation of orbital stability in the whole
orbital parameter space, we present in this paper a detailed
analysis on the dependence of orbital stability on the initial
eccentricity. We will explain the dynamical model, numerical
algorithm and analytical method used in this paper in Section 2.
The main results will be given in Section 3, where the dynamical
maps on the plane of $(a_0,e_0)$ are plotted and described. Then
in Section 4, through a frequency analysis method, we will discuss
the mechanisms that sculpture the dynamical maps. In Section 5,
the orbits of the 6 observed NTs will be checked and compared with
our previous results. Finally, a summation and a discussion will
be given in Section 6.

\section{Model and Method}
Basically, the dynamical model, the numerical method and the
analytical tools used in this paper are the same or very similar to
the ones in Paper I. We give a brief introduction below.

\subsection{Dynamical model and analyzing methods}
All the computations and analysis in this paper are under the
``outer Solar system model'' in which the massless fictitious
asteroids orbit around Neptune's triangular Lagrange points in the
gravitation field of the Sun and the four outer giant planets.
Since $L_4$ and $L_5$ are dynamically symmetrical to each other
\cite{nes02a,mar03a,zho09a}, we may investigate only one of them
without losing the generality. Below we will discuss only the
trailing point $L_5$.

To simulate the orbital evolution, we use the Lie-series integrator
\cite{han84}. An on-line low-pass filter is embedded in to remove
the short-term oscillations in the outputs \cite{mic95,mic02}. The
spectra of each orbit are then calculated from these filtered
outputs using the Fast Fourier Transform (FFT) method. The number of
peaks above a given level in a spectrum is counted. This so-called
``Spectral Number'' (SN) can be regarded as an indicator of the
regularity of an orbit \cite{mic95,mic02,sfm05}, and it is used to
construct the dynamical maps on the initial orbital parameter plane.
The frequencies obtained from the FFT procedure are also carefully
analyzed to determine the mechanisms that portrait the features in
the dynamical maps. For details of the above mentioned methods and
technologies, please refer to Paper I.

\subsection{Initial conditions and the libration center}
In order to get representative initial conditions for numerical
simulations, we follow a similar scheme as Nesvorn\'y and Dones
\shortcite{nes02a} did. The initial orbital elements of fictitious
asteroids are taken from a grid on the plane $(a_0, e_0)$, where
101 values of the semimajor axes $a_0$ are chosen uniformly
between 29.9 and 30.5\,AU, and the eccentricities $e_0$ vary up
from 0.0 to 0.5 with an increment of 0.005. For all orbits on the
grid, the initial inclinations $i_0$ are fixed at a given value.
The angular elements of the orbits, including the ascending node
$\Omega_0$, the perihelion argument $\omega_0$ and the mean
anomaly $M_0$, are arranged in such a way that the resonant angle
$\sigma$ ($\sigma= \lambda - \lambda_8$, the difference between
the mean longitudes of Neptune $\lambda_8$ and the fictitious
trojan $\lambda$) is at the center of tadpole orbit ($\sigma_c
\approx -60^\circ$). Specifically, we set $\Omega_0= \Omega_8,
\omega_0= \omega_8-60^\circ$ and $M_0= M_8+ \sigma_c+60^\circ$.

It is known that the libration center $\sigma_c$ changes with the
eccentricity and inclination. In the restricted three body model
consisting of the Sun, Neptune and the Trojan, the value of
$\sigma_c$ at a given eccentricity and inclination can be determined
by locating the extremes of the averaged Hamiltonian (energy) on the
$(\sigma, a)$\,-plane \cite{nes02b}. In the outer solar system model
adopted in this paper, because of the perturbations from other
planets, there could be no such a center with zero libration
amplitude of $\sigma$. However, we can still locate the position at
which $\sigma$ librates with the smallest amplitude. A series of
orbits are integrated and the one with minimum libration amplitude
is ``defined'' as the libration center.

\begin{figure}
\label{fig01}
 \vspace{7.5cm}
 \includegraphics{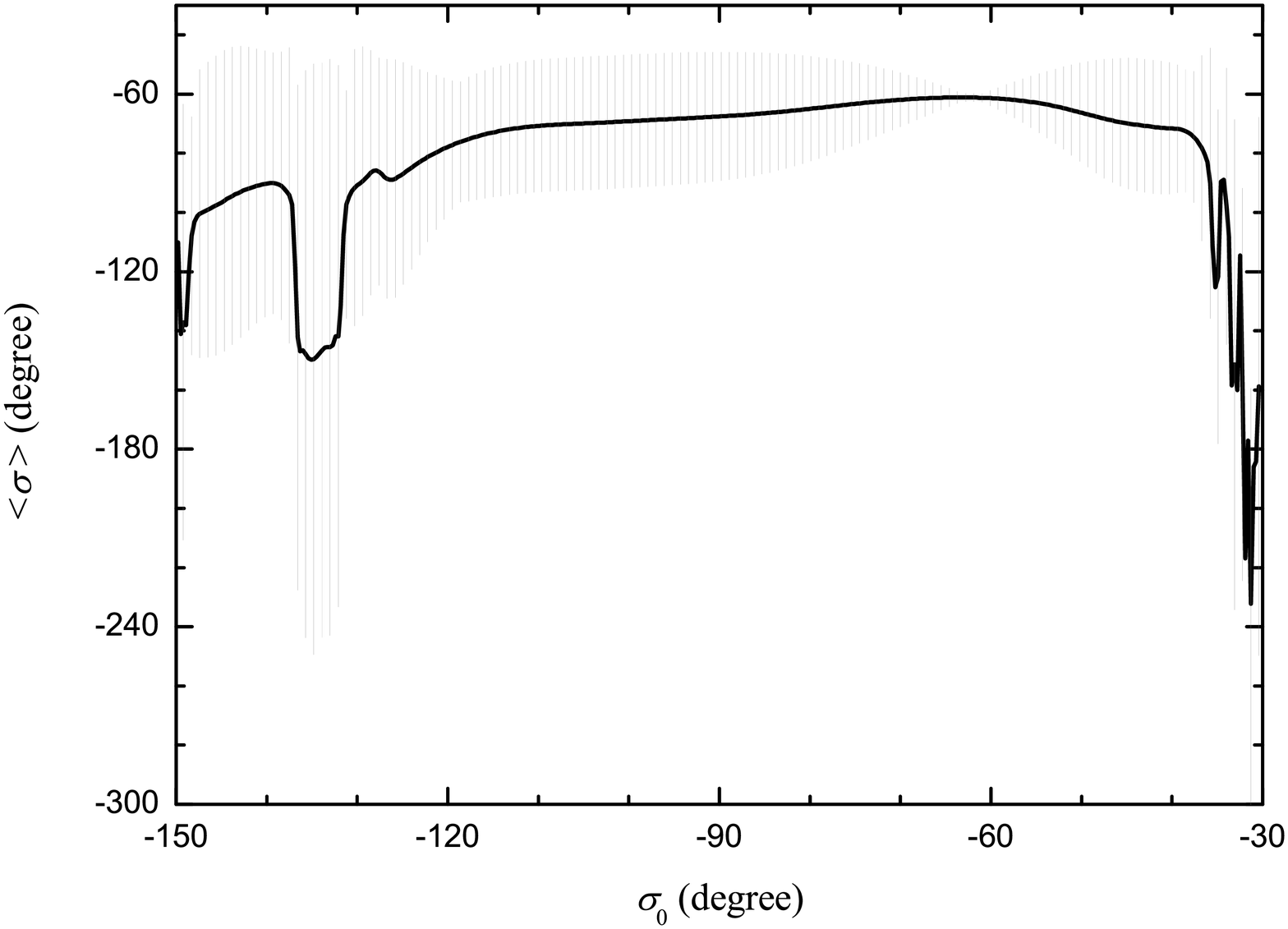}
 \caption{The mean value (solid curve) and the libration amplitude (error
 bars) of $\sigma$ versus initial $\sigma_0$. These 101 orbits have the same
 orbital elements except their mean anomalies differ from each other so that
 they have different initial resonant argument $\sigma_0$. The initial
 eccentricity and inclination of these orbits are $e_0=0.1, i_0=0^\circ$.
 The minimum of libration amplitude
 ($\Delta\sigma=6.97^\circ$) is at $\sigma_c=-64^\circ$.}
\end{figure}

In this series of orbits, the initial semimajor axes are chosen as
the same value as Neptune's, $a_0=a_8$ and $e_0, i_0$ are given
the specific values. The initial angular elements are given the
representative value as mentioned above, $\omega_0= \omega_8-
60^\circ, \Omega_0=\Omega_8, M_0=M_8+60^\circ+\sigma$. We test
$\sigma$ in the range $(-100^\circ, -50^\circ)$. The orbits are
then integrated up to $\sim 3.3\times 10^5$\,yr, which is about 4
times longer than the typical long-term variation period of
$\sigma$ \cite{mur99}. We determine the ``practical'' libration
center from the behaviour of the mean value and the librating
amplitude of $\sigma$. An example with initial eccentricity
$e_0=0.1$ and inclination $i_0=0^\circ$ is shown in Fig.\,1.

It is interesting to notice in Fig.\,1 that around the libration
center the libration amplitude increases with the difference of the
initial $\sigma_0$ from the center $\sigma_c$ and both of the mean
value and oscillation amplitude change continuously, reflecting the
regularity of motion around the libration center. When the initial
$\sigma_0$ is far away from the center, e.g. $\sigma_0<-90^\circ$,
the libration amplitude becomes large and the mean value changes
randomly, indicating the chaotic character of motion.

Varying $e_0, i_0$ and repeating the above calculations, we find
the variation of libration center with respect to $e_0$ and $i_0$.
A summation is plotted in Fig.\,2 for orbits with initial
conditions of $e_0=0 - 0.30$ and $i_0=0^\circ - 60^\circ$. We
ignore higher eccentricity and inclination because the highly
eccentric and/or inclined orbits are unstable and therefore are of
less interest. In fact, at very high eccentricity, the libration
center may ``bifurcate'' to two centers. But this phenomenon is
out of the range of current paper.

\begin{figure}
\label{fig02}
 \vspace{7.5cm}
 \includegraphics{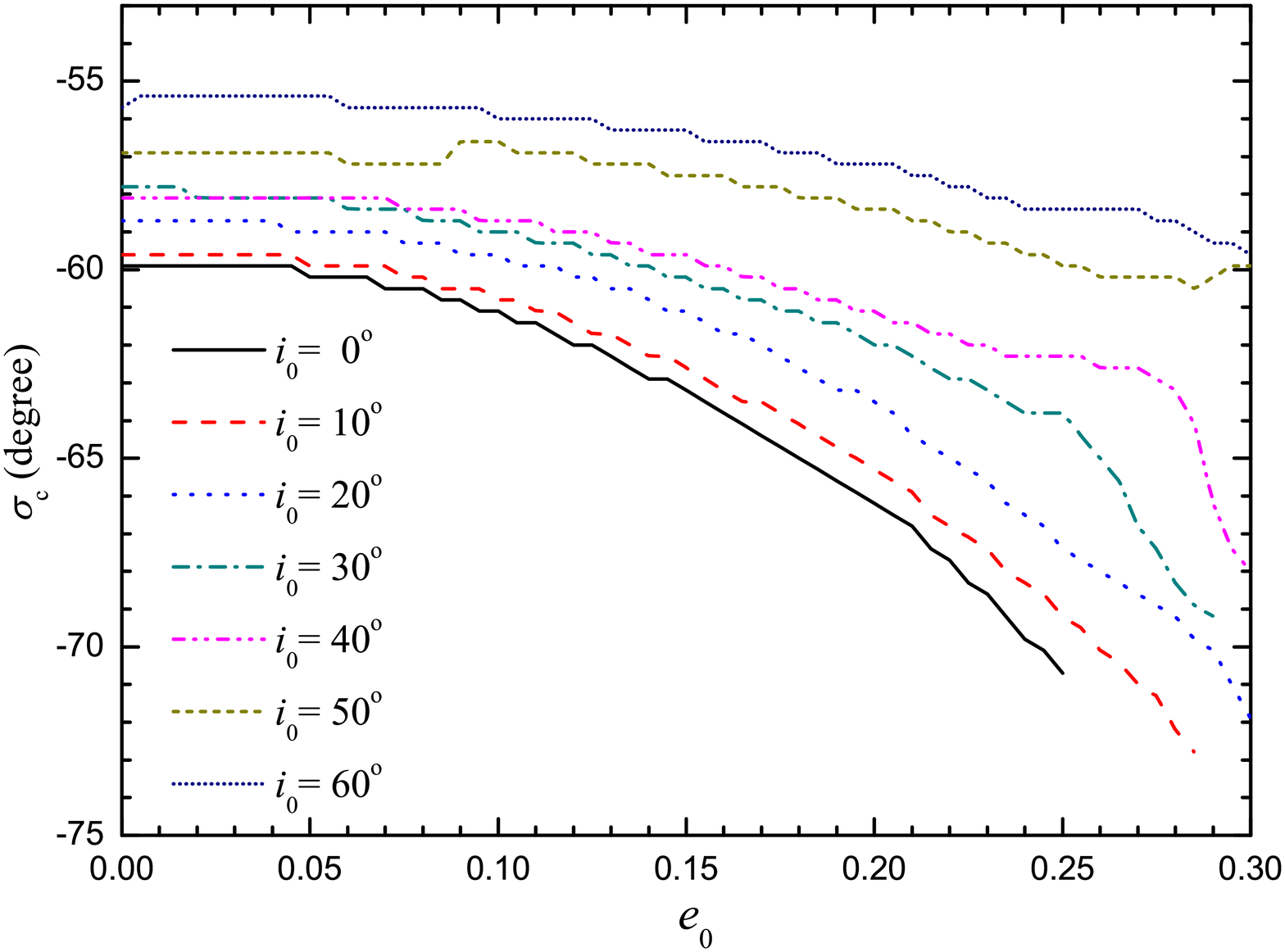}
 \caption{The variation of libration center ($\sigma_c$) with respect to
 eccentricity. From bottom to top, the curve represents the case for initial
 inclination from $0^\circ$ to $60^\circ$, respectively.}
\end{figure}

As shown in Fig.\,2, the libration center of the tadpole orbit
shifts farther away from Neptune as the eccentricity of the orbit
increases. For the coplanar orbit ($i_0=0^\circ$), the libration
center is exactly at the equilateral triangular point with
$\sigma_c=-60^\circ$ when the eccentricity is 0, and it changes
continuously to $\sigma_c\sim-71^\circ$ when the eccentricity
increases to 0.25. The $\sigma_c$ also changes a little with respect
to the inclination, and compared with coplanar orbits, the variation
of $\sigma_c$ with eccentricity for highly inclined orbits becomes
smaller.

Only after knowing the libration center $\sigma_c(e_0,i_0)$ at
different eccentricity ($e_0$) and inclination ($i_0$), we can
select the proper initial conditions to construct the representative
dynamical maps. Namely, for a given inclination $i_0$, a dynamical
map on the $(a_0,e_0)$-\,plane is constructed by analyzing the
stability of each orbit on this plane and with other initial angular
elements given by $\Omega_0= \Omega_8, \omega_0= \omega_8-60^\circ,
M_0= M_8+60^\circ+ \sigma_c(e_0,i_0)$.

\section{Dynamical maps}
We have investigated in Paper I the dependence of orbital
stability of NT on the inclination. In this paper, we focus on the
dependence of stability on the eccentricity. For this sake, the
initial inclinations $i_0$ is fixed at given values and we study
the orbits of the fictitious trojans initialized on the grid on
the $(a_0,e_0)$-\,plane with other orbital elements determined in
a way described in last section. Using the ``spectra number'' (SN)
as the indicator of the regularity of an orbit as in Paper I, we
construct dynamical maps on the $(a_0,e_0)-$\,plane.

We present in Fig.\,3 the dynamical maps for $i_0 = 0^\circ,
10^\circ, 20^\circ, 30^\circ, 40^\circ$ and $i_0=50^\circ$. The
color in these dynamical maps indicates the orbital regularity.
The green represents the most regular orbit, the red is on the
edge of chaotic motion and the white indicates that the orbit does
not survive on the trojan-like orbit in our numerical simulation
(which lasts about 34\,Myr). According to our experiences in Paper
I, orbits with SN$>50$ (color blue) are not to be expected to
survive in the Solar system age. Surely it is impossible and also
unnecessary to calculate each slice at every possible initial
inclination. These slices in Fig.\,3 are enough to show the
variation of dynamical map with respect to the initial
inclination. We also discard those cases in which most of the
trojan orbits are unstable and therefore less interested. For
instance, due to the depletion effects arising from the $\nu_8$
secular resonance and Kozai resonance, the motion of trojans with
$i_0 \sim 45^\circ$ and $i_0>60^\circ$ are unstable (see Paper I),
so that the dynamical maps are plain and not shown here. However,
for the sake of providing more information about the trojan motion
at other inclinations, other three dynamical maps for
$i_0=5^\circ, 35^\circ$ and $55^\circ$ are used in
Figs.\,5,\,6\,\&\,7 as examples to present our frequency analysis
in next section.

\begin{figure*}
\label{fig03}
 \vspace{22.5 cm}
 \includegraphics{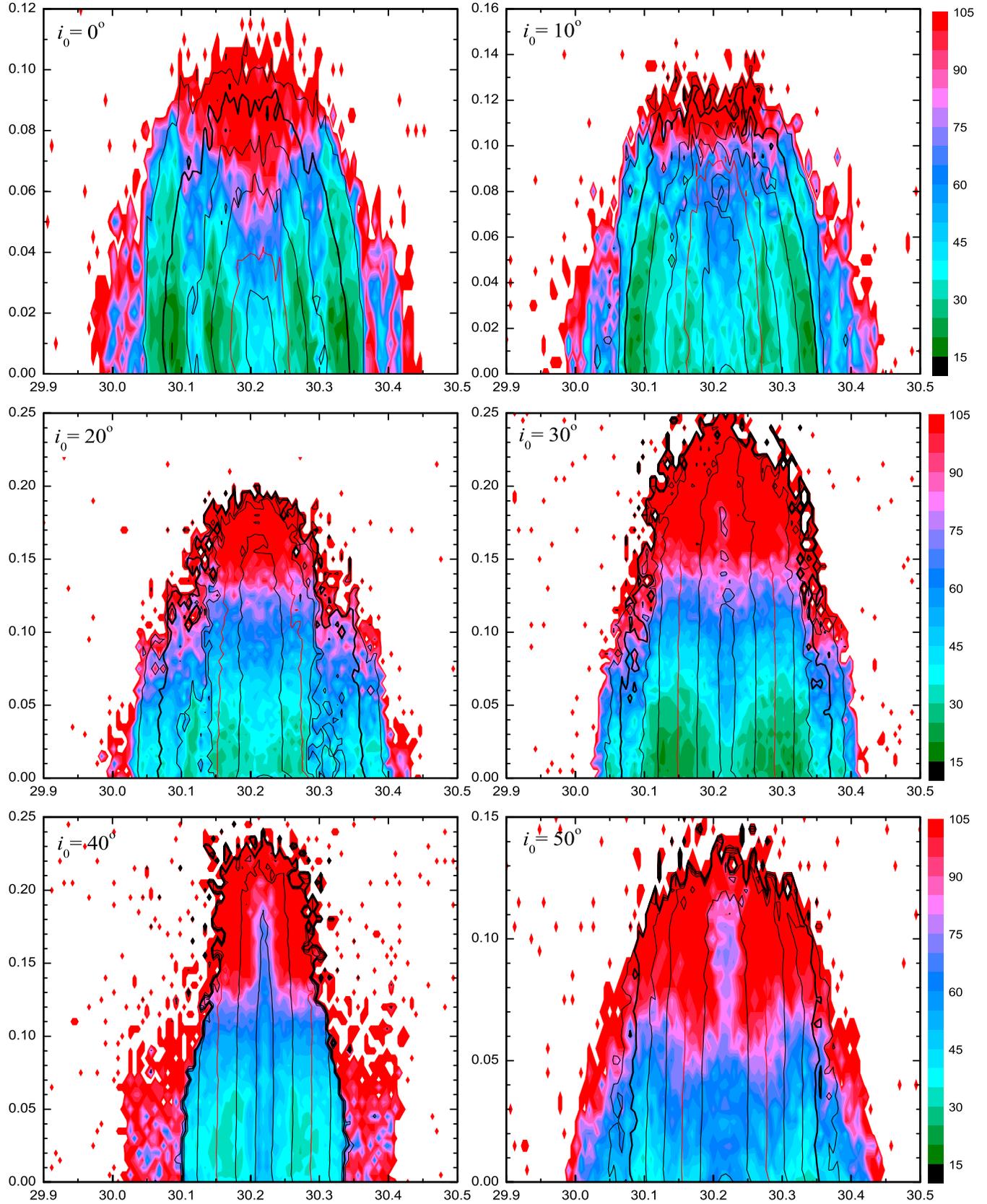}
 \caption{Dynamical maps on $(a_0,e_0)-$\,plane, for initial inclinations
 $0^\circ, 10^\circ, 20^\circ, 30^\circ, 40^\circ$ and $50^\circ$ (left to
 right, top to bottom). The abscissae and ordinates are initial semimajor
 axes and eccentricities. Note the ordinates may have different scales. The
 contour curves represent the libration amplitude of resonant angle $\sigma=
 \lambda-\lambda_8$. From inside out are contours for $\Delta\sigma= 10^\circ$
 to $\Delta\sigma= 70^\circ$ with an increment of $10^\circ$. Particularly, the
 red curves are for $\Delta\sigma= 30^\circ$ and the thick ones represent $\Delta\sigma=
 60^\circ$.}
\end{figure*}

The first conclusion drawn from these dynamical maps may be that the
most stable motion (color green) does not happen at any inclination,
but only at $i_0= 0^\circ, 5^\circ, 10^\circ, 30^\circ, 35^\circ$
and $55^\circ$. This is consistent with the finding in Paper I of
three most stable regions in inclination: Region A $(i_0 = 0^\circ -
12^\circ)$, B $(i_0= 22^\circ - 36^\circ)$ and C $(i_0 = 51^\circ -
59^\circ)$.

A second conclusion is that the initial eccentricity $e_0$ for
stable orbits must be small. Although for some initial
inclinations (e.g. $i_0= 30^\circ, 35^\circ$) some orbits with
$e_0$ as large as 0.25 survive in our integrations and may survive
even for much longer time, their SN's (color red) indicate that
most probably they will not remain on the trojan-like orbits in
the Solar system age. These orbits with high $e_0$ may suffer
perturbations through resonances whose effects appear
significantly only in a timespan much longer than our integration
time. For the most stable orbits, the largest initial
eccentricities are around 0.08 in Region A, 0.07 in Region B and
0.02 in Region C. However, if we use a lenient criterion, namely,
regarding all the orbits with SN$ < 50$ as survival ones (in the
Solar system age), the permitted eccentricity extends to 0.10,
0.12 and 0.04 in Region A, B and C respectively. Out of these
three most stable regions in inclination, there are also some
narrow areas in which the motion is stable. For example, on the
slice of $i_0= 20^\circ$ in Fig.\,3, we see a few stable orbits
with $e_0 < 0.05$.

The contour of libration amplitude of the resonant angle
$\Delta\sigma = \sigma_{\rm max}- \sigma_{\rm min}$ is also
calculated and over-plotted on the dynamical maps. Clearly the
relation between $\Delta\sigma$ and the deviation of semimajor axis
from its mean value $a_0-a_c$ \cite{erd88,mil93,zho09a} is obeyed.
The $\Delta\sigma$ increases monotonically with the distance of
initial semimajor axis $a_0$ from its value at libration center ($
a_c \sim 30.22$\,AU). The maximum $\Delta\sigma$ is around
$\Delta\sigma \sim 70^\circ$ for stable orbits with low (Region A)
and medium inclination (Region B), but shrinks a little for high
inclination (Region C) to $\Delta\sim 50^\circ$. At given $i_0$ and
$a_0$, the libration amplitude increases only very little as the
initial eccentricity increases. At high eccentricity, the
$\Delta\sigma$ may still be small, but the large SN implies a
chaotic motion. Therefore, the small libration amplitude manifested
in our integration time may increase in a distant future. We also
note that there is a lack of small $\Delta\sigma$ at some
inclination values. For example, at $i_0=0^\circ, 10^\circ$ and
$50^\circ$ all trojan orbits have $\Delta\sigma > 10^\circ$. Last
but not least, thanks to the calculation of libration center
$\sigma_c$ and proper selection of initial conditions, both the
dynamical maps and the libration amplitude contours have the
bilaterally symmetrical features. Otherwise, if we just simply set
$\sigma_c=-60^\circ$, these figures would be distorted.

Contrary to the profiles of the libration amplitude contours that
are more or less plain, the dynamical maps bear plenty of fine
structures. Most distinctly, there are vertical gaps of irregular
motion on slices of small inclination ($i_0=0^\circ, 5^\circ,
10^\circ$), and a ``spike'' of relatively regular orbits with small
$\Delta\sigma$ can be clearly seen on slices of large inclination
($i_0=40^\circ, 50^\circ, 55^\circ$). On one hand, these fine
structures reveal the peculiar behaviour of orbits in these regions;
on the other hand, they supply hints of the mechanisms taking
effects and portraying the dynamical maps in the corresponding
parameter space. We will try in next section to locate different
kinds of resonances on the $(a_0,e_0)$-\,plane and compare them with
the fine structures in the dynamical maps.

We would like to notify that the dynamical maps are plotted based
on the osculating orbital elements in our papers. As we have shown
in Paper I, the dynamical map would change if a different epoch
was chosen. In a different epoch, the features of the dynamical
map are very well kept, and the change is nothing more than a
shift in the semimajor axis. If the proper libration amplitude and
proper eccentricity, instead of the osculating semimajor axis and
osculating eccentricity, are used to define the representative
plane for the dynamical map, we may obtain the epoch-free
dynamical maps as in \cite{mar03a,mar03b,dvo07}. But the
osculating elements have the merit of being able to compare with
the observing data directly. To do this, all we need is to
transform the observing data to the corresponding epoch, as we
will do in Section 5 of this paper.

\begin{figure}
\label{fig04}
 \vspace{7.5cm}
 \includegraphics{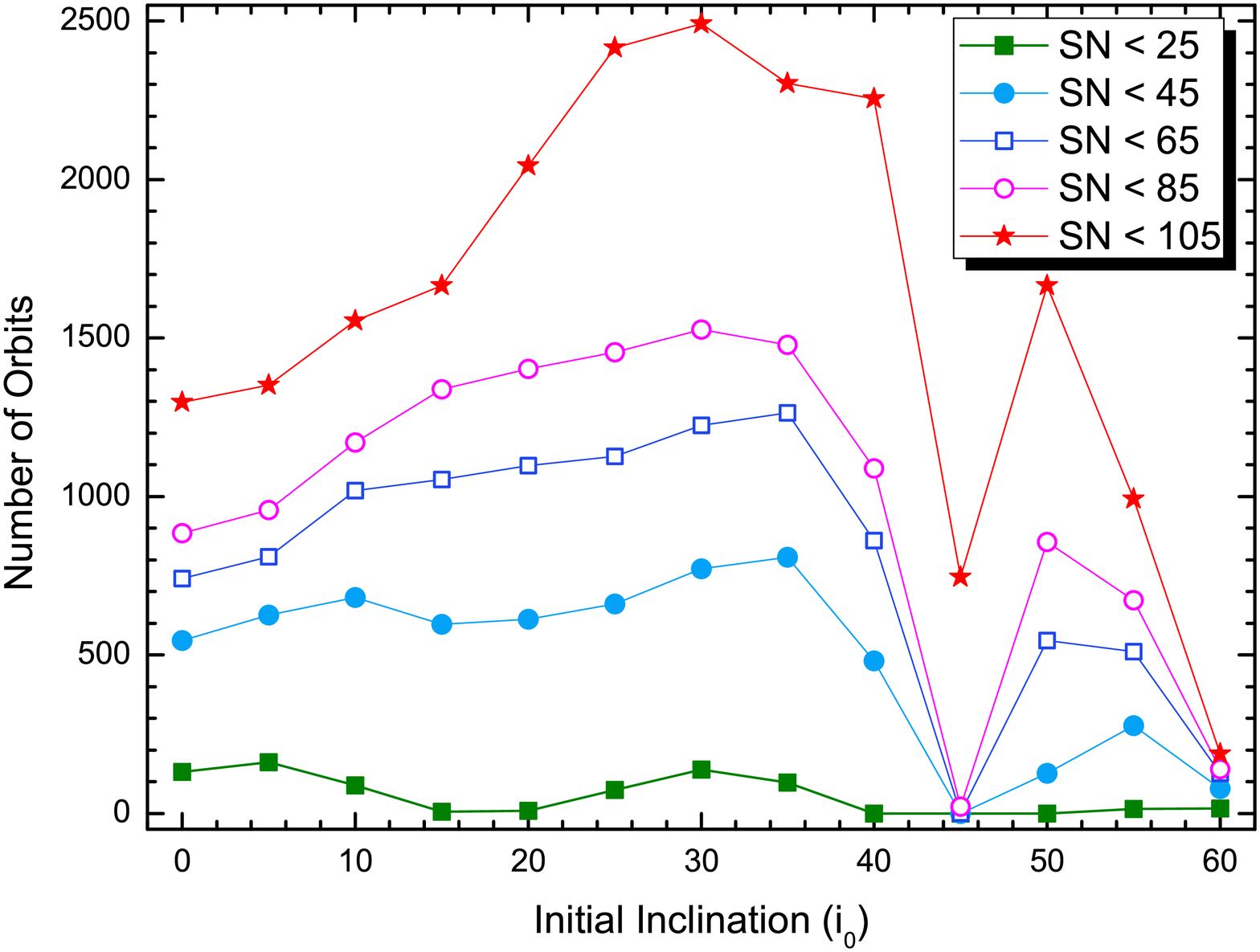}
 \caption{The number of regular orbits versus the initial inclination. From
 bottom up, the solid squares, solid circles, open squares, open circles and
 stars indicate the numbers of orbits whose SN's are smaller than 25, 45, 65,
 85 and 105. Remember all those orbits with original ${\rm SN} >100$ and
 $29.9\,{\rm au}<\bar{a}<$ 30.5\,au are assigned a new SN of 100 (please refer
 to Paper I for details).}
\end{figure}

We have calculated many other dynamical maps with different
initial inclinations, of which three maps for $i_0=5^\circ,
35^\circ$ and $i_0=55^\circ$ are going to be shown in next
section. Instead of presenting all of them here, we count the
number of regular orbits on the dynamical map of each selected
inclination value. Since for each map we have calculated 5151
orbits on a $101\times 51$ grid of $(a_0, e_0)$, the number of
orbits with the SN smaller than a definite value is an indicator
of the relative regularity of orbits. The results are summarized
in Fig.\,4, where we show the numbers of orbits with SN smaller
than 25, 45, 65, 85 and 105. According to our rule of assigning
SN, all asteroids that escaped from the1:1 MMR (judged by their
mean semimajor axes during the orbital integrations) have been
designated an SN of 110. Thus the stars in Fig.\,4 imply the
surviving probability of artificial Trojans out of those totally
5151 ones in our integration time.

The survival number of artificial Trojans (stars in Fig.\,4)
increases with the initial inclination before it reaches its
maxima at $i_0=30^\circ$. Then it drops down to a minimum at $i_0
= 45^\circ$ due to the depletion by the $\nu_8$ secular resonance,
and another maximum is reached at $i_0 = 50^\circ$. It ends at
higher inclination where the Kozai resonance dominates (Paper I).
If adopting the survival probability from our integrations as the
stability indicator, we would conjecture that more NTs could be
found on high-inclined orbits with inclination around $30^\circ$
and $50^\circ$. But if we focus on orbits with ${\rm SN}<25$
and/or ${\rm SN}<45$, we would argue that NTs may mainly
concentrate in a low-inclination region around $i_0 \sim 5^\circ$
and a medium-inclination region around $i_0 \sim 30^\circ$. This
has been partly proven to be true by the observations, namely,
among the six observed NTs, 4 are found in the former region and
the rest 2 in the latter. In addition, judging from Fig.\,4, a
high-inclination region around $i_0 \sim 55^\circ$ also could host
some NTs. However, we would note that the stability of the
high-inclined orbits may be determined by the dynamical analysis
under current configuration of the Solar system, but the existence
of real NTs on the specific orbits depends also on the evolution
of planetary orbits in the early stage of the Solar system
\cite{kor04,lij07,nes09}.

\section{Frequency analysis}
\subsection{Basic ideas}
As we have mentioned above, some secondary resonances and secular
resonances are responsible for the forming of the fine structures
in the dynamical maps. To understand the dynamics of NTs clearly,
we need to find the resonances associated with these structures.
One natural and simple way to determine a resonance is to check
the behaviour of the corresponding critical resonant angle, whose
librating character is an evident sign of being inside a
resonance. But, as mentioned in Robutel \& Gabern
\shortcite{rob06}, this method is valid only for low order MMRs or
secular resonances of second order. It doesn't work when the
resonance involves more than two degrees of freedom or when it is
of high order. In fact, a typical power spectrum of a complex
motion generally is a composition of an amount of terms with
comparable amplitudes and different frequencies, this means, the
regular variation of a critical angle of a high-order resonance
may be enshrouded by other high-amplitude terms even when the
motion is deep inside the resonance. Of course, the proper
critical angle can still be well defined if we can remove the
irrelevant high-amplitude terms by (for example) reducing the
corresponding Hamiltonian to a normal form \cite{rob06}. To define
such critical angles is out of the scope of this paper, so that we
will not try to check whether the resonant angle is librating or
circulating when we discuss a resonance. Instead, two techniques
are applied in this paper. One is the ``dynamical spectrum'' and
the other is the ``empirical formulation''. These two methods have
been used in Paper I, and we will give a very brief introduction
below.

For each NT orbit in our simulation, we calculated the power
spectra of $a\cos\sigma, e\cos\varpi$ and $i\cos\Omega$. These
spectra generally are complex compositions of the forced
frequencies, free frequencies, their harmonics and combinations of
above terms. From a spectrum, the most significant peaks are
selected and their frequencies are recorded.

We vary the orbital parameter and calculate these leading
frequencies at each parameter value. By plotting all these
frequencies on one figure (called ``dynamical spectrum'' as in Paper
I), we may find how the dominating frequencies of the motion change
with respect to the parameter. Knowing the secular frequencies of
the outer solar system, i.e. the apsidal and nodal precession
frequencies of Jupiter, Saturn, Uranus and Neptune ($g_5, g_6, g_7,
g_8$ and $s_5, s_6, s_7, s_8$), the forced frequencies can be easily
recognized. Since the proper frequencies of the motion generally
change continuously with the orbital parameter, they can also be
recognized. In a dynamical spectrum, along with the varying
parameter, a continuously varying frequency may meet and cross
another frequency, where a resonance associated with these
frequencies happens.

Thus a dynamical spectrum can be used to detect and locate the
resonances in the motion, and more importantly, it helps us
determine the proper libration and precession frequencies of the
motion. The proper frequencies of the NT's motion include the
libration frequency of the critical angle of the 1:1 MMR,
$f_{\sigma}$, the frequencies of the apsidal precession $g$ and of
the nodal precession $s$. They can be derived from the dynamical
spectra of the variables $a\cos\sigma, e\cos\varpi$ and
$i\cos\Omega$ respectively. With the data of the proper
frequencies at different parameter values in hand, we can compute
the empirical formulations for the proper frequencies with respect
to orbital parameters (e.g. the semimajor axis, inclination and
eccentricity) on the whole parameter space. Finally, with these
empirical formulations and those already-known frequencies, we may
calculate all kinds of equations of frequencies that define the
resonances in the parameter space.

Since the dynamical spectra are generally overloaded with details
(see for example Figs. 9\,\&\,10 in Paper I), we would rather not
show and describe them here. Instead, we will give directly the
empirical formulae of $f_\sigma, g, s$ derived from the dynamical
spectra, and then with the formulae we figure out the resonances
revealed by the dynamical spectra and manifested in the dynamical
maps.

To obtain a global view of the dynamics of NTs at different
inclinations, we analyze three typical dynamical maps on the
$(a_0,e_0)-$\,plane with $i_0=5^\circ, 35^\circ$ and $55^\circ$,
respectively. These selected slices are inside the most stable
regions Region A, B and C.

\subsection{For $i_0=5^\circ$}
Adopting the similar quadratic formula \cite{mil94,mar03b} as in
Paper I, and setting $\delta=a_0-30.215$, we derive from our
calculations the empirical expressions of $f_\sigma, g, s$ on the
$(a_0,e_0)-$\,plane for the slice $i_0=5^\circ$, which are:
\[
 f_\sigma [10^{-4}\,2\pi{\rm /yr}]=1.1350-2.1783\delta^2-0.046311e^2
\]
\begin{equation}
\\+2.0566\delta^4 -27.500e^4+28.700\delta^2e^2,
\end{equation}
\[
g [10^{-6}\,2\pi{\rm /yr}]=1.5783+6.9896\delta^2+3.3556e^2
\]
\begin{equation}
 \\-72.941\delta^4 +174.70e^4-110.97\delta^2e^2,
\end{equation}
\[
 s [10^{-7}\,2\pi{\rm /yr}]=5.1205+21.435\delta^2+19.102e^2
\]
\begin{equation}
\\-54.217\delta^4+3567.4e^4-416.73\delta^2e^2.
\end{equation}
Similar calculations are performed for slices of $i_0=35^\circ,
55^\circ$ and will be presented below. Using these formulae, all
the locations of different kinds of commensurabilities between
frequencies can be computed. However, before doing this, let us
recall what kinds of resonances may happen in NTs' motion.

In the study of Jupiter Trojans, Robutel \& Gabern \shortcite{rob06}
defined four families of resonances as follow:
\begin{equation}
\begin{array}{ll}
{\rm I}:   & pf_\sigma-n_5+qg+q_5g_5+q_6g_6=0; \\
{\rm II}:  & f_\sigma+5f_{5:2}+pg+p_5g_5+p_6g_6=0; \\
{\rm III}: & qs+q_6s_6+p_5g_5+p_6g_6=0; \\
{\rm IV}:  & pg+f_{5:2}+p_5g_5+p_6g_6=0.
\end{array}
\end{equation}
In above equations, $n_5$ is the mean motion of Jupiter, $f_{5:2}$
is the frequency of the quasi 5:2 resonance between Jupiter and
Saturn (the Great Inequality), and all the integers
$p,q,p_5,p_6,q_5$ and $q_6$ should be chosen in a way such that the
d'Alembert rule is fulfilled.

It is reasonable to postulate that similar resonances are involved
in the case for NTs' motion. Among the already-known frequencies,
except for those secular frequencies $g_5, g_6, g_7, g_8$ and
$s_5, s_6, s_7, s_8$, we find the frequency of the quasi 2:1 MMR
between Neptune and Uranus, that is, the frequency of the angle
$2\lambda_8-\lambda_7$ (denoted by $f_{2:1}$ hereafter), plays an
important role in the motion of NT. On the contrary, the
perturbing effects from the Great Inequality is (relatively)
ignorable, due to its far distance and perhaps also due to its
high frequency ($\sim 10^{-3}\,2\pi/{\rm yr}$).

In addition, we notice that the mean motion of Neptune $n_8
(\approx 6.068\times 10^{-3}\,2\pi/{\rm yr})$ is much higher than
the libration frequency $f_\sigma (\sim 1\times 10^{-4}\,2\pi/{\rm
yr})$. This means, if a resonance similar to Family-I happened,
the integer $p$ would be as large as $\sim 60$. Moreover,
$f_{2:1}=2.36064\times 10^{-4}\,2\pi/{\rm yr}$ is at least 2
orders of magnitude larger than the precession rate $g, s$ (and
also other precession frequencies related to planets). Therefore,
we may ignore those resonances as Family-I and Family-IV in NT's
motion, because their order must be quite high. Instead, only two
types of resonances are discussed in this paper:
\begin{equation}
\begin{array}{ll}
{\rm S}: & pg+qs+\sum_{j=5}^{8}(p_jg_j+q_js_j)=0, \\
{\rm C}: & hf_\sigma+kf_{2:1}+pg+qs+\sum_{j=5}^{8}(p_jg_j+q_js_j)=0,
\end{array}
\end{equation}
where $h,k,p,q,p_j,q_j$ are integers. The ``S'' stands for
``Secular'', indicating they are secular resonances in the usual
sense. While the ``C'' for ``Combined'' indicates that these
resonances are combinations of mean motions and secular
frequencies. The d'Alembert rule for S-type resonance is
$p+q+\sum_{j=5}^{8}(p_j+q_j)=0$, and it's
$k+p+q+\sum_{j=5}^{8}(p_j+q_j)=0$ for C-type. In both equations,
$(q+\sum_{j=5}^{8}q_j)$ must be even so that the symmetry of
inclination with respect to the reference plane can be guaranteed
\cite{mur99}.

By checking carefully the details in the dynamical spectra and using
the empirical expressions in Eqs.\,(1)-(3), we recognize some
important resonances taking effects on the slice of $i_0=5^\circ$,
whose locations are calculated and then plotted over the
corresponding dynamical map as shown in Fig.\,5.

\begin{figure}
\label{fig05}
 \vspace{7.5cm}
 \includegraphics{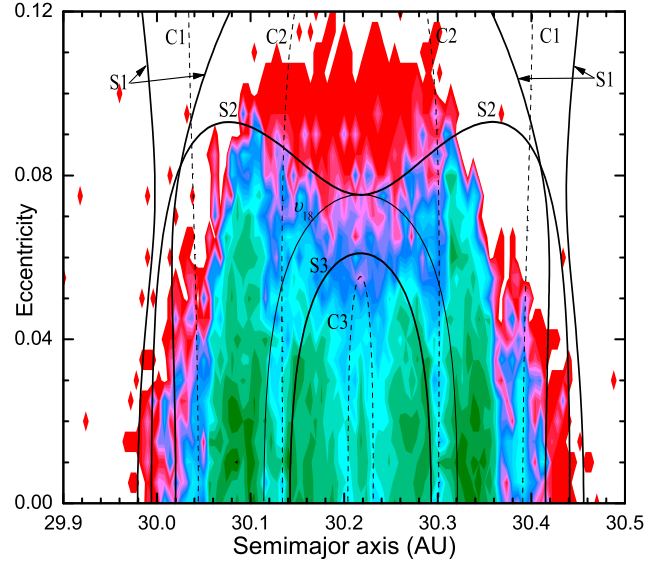}
 \caption{The locations of resonances plotted over the dynamical map for the
 case with initial inclination $i_0=5^\circ$. The C-type resonances are indicated
 by dashed curves while the solid curves stand for the S-type resonances,
 particularly, the secular resonance $\nu_{18}$ is plotted by the thin solid curve.
 See text to find the meaning of labels C1, C2, S1, S2 and S3.}
\end{figure}

The most distinguishable structure in the dynamical map of
$i_0=5^\circ$ (and also other slices at low inclination, e.g.
$i_0=0^\circ, 10^\circ$ in Fig.\,3) is the blue vertical
``stripes'' indicating less stable motion. We find that they arise
from the effects of some C-type resonances, among which three
major ones are shown in Fig.\,5. The dashed curves labeled by C1,
C2 and C3 are locations of the following resonances:
\begin{equation}
\begin{array}{ll}
{\rm C1}: & 2f_\sigma-f_{2:1}+g_6=0, \\
{\rm C2}: & 4f_\sigma-2f_{2:1}+g_6+g_7=0, \\
{\rm C3}: & 6f_\sigma-3f_{2:1}+g_5+g_6+g_7=0.
\end{array}
\end{equation}
Apparently, their locations match the positions of vertical
structures very well. We need to stress here that each resonance
listed above (and below too) is just a representative of a bunch
of resonances, characterized by higher order combinations of
integers $p, q, p_j, q_j$ in Eq.\,(5). For example, beside the
resonance C1 that reaches the abscissa at $a_0=30.044$\,au and
$30.392$\,au, we can easily find another two resonances as ${\rm
C1^\prime:}\, 2f_\sigma - f_{2:1} - g + g_8 + g_6=0$ and ${\rm
C1^{\prime\prime}:}\, 2f_\sigma - f_{2:1} +g-g_8+g_6=0$. And in
fact, the resonance C1 itself is favourable for the stability of
Trojan orbit. As we will see in next Section, the asteroid 2001
QR322 is deep inside such a resonance and its orbit is stable over
the Solar system age. On the contrary, the resonances ${\rm
C1^\prime}$ and ${\rm C1^{\prime\prime}}$ are responsible for the
unstable gaps in both sides of resonance C1, especially the
apparent ones around $a_0=30.050$\,au and $30.375$\,au
approximately.

Another mechanism that may contribute to the formation of the
vertical structures is the secular nodal resonance $\nu_{18}$
defined by $s-s_8=0$. Its location is also plotted in Fig.\,5, as
well as other three S-type resonances:
\begin{equation}
\begin{array}{ll}
{\rm S1}: & 3g-g_7-g_8=0, \\
{\rm S2}: & g-2s+g_8-2s_8+2s_5=0, \\
{\rm S3}: & 3s-2g_8-s_8=0.
\end{array}
\end{equation}

We notice inconsistent statements about the resonance $\nu_{18}$
in the literatures. Nesvorn\'y and Jones \shortcite{nes02a} found
in the dynamical map a less stable region with libration amplitude
between $40^\circ$ and $50^\circ$ (Fig.\,2(d) therein), and they
argued that this arose from the effects of the $\nu_{18}$
resonance. While Brasser et al. \shortcite{bra04} showed that the
clones of asteroid 2001 QR322 ``remained deep inside'' the
$\nu_{18}$ resonance are the most stable ones (``have the longest
e-folding time''). Marzari and Scholl \shortcite{mar03a} argued
that this secular resonance were ``not strong enough to cause a
short term instability''. Our results support the same conclusion.

In Fig.\,5, no apparent details in the dynamical map can be found
around the location of $\nu_{18}$ resonance. In fact, we found in
our simulations that nearly all orbits with low inclination ($i_0
\le 5^\circ$) and not too large libration amplitude are involved
in the $\nu_{18}$ resonance, but this resonance does not show any
prominent effects in protecting or destroying their orbital
stabilities. The less stable region mentioned in Nesvorn\'y and
Jones \shortcite{nes02a} in fact corresponds to the blue stripes
at $a_0=30.12$\,au and $30.30$\,au in the dynamical map of
$i_0=0^\circ$ in Fig.\,3, and it is clearly related to the C2
resonance in Eq.\,(6), but not the $\nu_{18}$ resonance.

Other secular resonances, e.g. the S1, S2 and S3 in Eq.\,(7),
however may have more evident dynamical effects. The resonances S1
and S2 define the edge of the survival region, and the curvature of
S2 in the eccentricity range of $0.07 - 0.09$ matches the outline of
the stable region. The border between the stable region and chaotic
region makes a V-shape blue area from $e_0=0.05$ to $e_0=0.09$, in
which several crossovers or close encounters between resonances S2,
S3, C2, C3 and $\nu_{18}$ happen. Thus it seems reasonable to argue
that the border is made of a net of these resonances.

Another observing from Fig.\,5 and other dynamical maps at different
initial inclinations in Fig.\,3 is that the C1-type resonance with
$h=2, k=-1$ in Eq.\,(5) has quite promising dynamical effects in all
initial inclination values, but the influences from C2 and C3
decreases as $i_0$ increases. The blue gaps corresponding to C2 and
C3 become more and more vague as $i_0$ gets bigger. We will see this
trend clearly in next example of $i_0=35^\circ$.

\subsection{For $i_0=35^\circ$}
Similarly, setting $\delta=a_0-30.220$, we obtain the best fits of
the proper frequencies formulae for the slice $i_0=35^\circ$:
\[
 f_\sigma [10^{-4}\,2\pi{\rm /yr}]=1.0109-1.5923\delta^2+0.45999e^2
\]
\begin{equation}
\\-2.1860\delta^4 -29.100e^4-5.1370\delta^2e^2,
\end{equation}
\[
g [10^{-7}\,2\pi{\rm /yr}]=8.3905-4.9257\delta^2-6.113e^2
\]
\begin{equation}
 \\-241.09\delta^4 -75.306e^4-347.56\delta^2e^2,
\end{equation}
\[
 s [10^{-7}\,2\pi{\rm /yr}]=2.8989+15.920\delta^2+7.1909e^2
\]
\begin{equation}
\\-62.502\delta^4+161.75e^4+368.27\delta^2e^2.
\end{equation}
Then by carefully studying the dynamical spectra, we detect some
resonances involved in the motion. Using the above formulae, we
calculate their locations and plot them over the corresponding
dynamical map in Fig.\,6.

\begin{figure}
\label{fig06}
 \vspace{7.5cm}
 \includegraphics{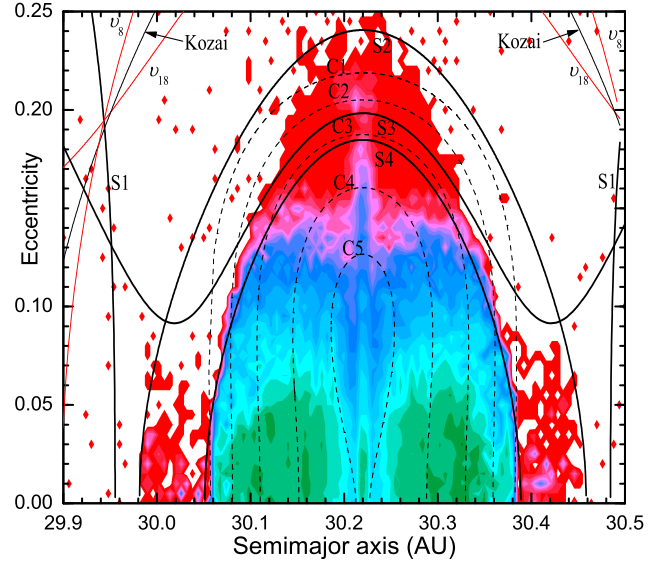}
 \caption{The same as Fig.\,5 but for $i_0=35^\circ$. Except the resonances
 labeled by C1-C5 (dashed curves) and S1-S4 (solid curves. See text for their
 meanings), the locations of $\nu_8, \nu_{18}$
 resonances (solid curves in red) and Kozai resonance (thin solid curve in black)
 have also been plotted.}
\end{figure}

As in the previous case, we find in our calculations some C-type
resonances involved in the motion of orbits with $i_0=35^\circ$, of
which several typical ones are listed below and plotted in Fig.\,6.
\begin{equation}
\begin{array}{ll}
{\rm C1}: & 2f_\sigma-f_{2:1}-g+2g_6=0, \\
{\rm C2}: & 2f_\sigma-f_{2:1}-s+2s_6-s_8+g_8=0, \\
{\rm C3}: & 2f_\sigma-f_{2:1}+g+2g_6-2g_5=0,\\
{\rm C4}: & 2f_\sigma-f_{2:1}+2g+2g_6-3g_5=0, \\
{\rm C5}: & 4f_\sigma-2f_{2:1}+3g_6+g_5-2g_8=0.
\end{array}
\end{equation}
The low-order C-type resonance like the C1 defined in Eq.\,(6)
does not appear in the region around $i_0=35^\circ$.
Among the resonances in Eq.\,(11), only the one with the lowest
order (C1) has the distinguishable dynamical effects, shaping the
sharp edge of stable region at low eccentricity ($e_0\le 0.05$).
The C-type resonance with $h=4, k=-2$ in Eq.\,(5), here C5 in
Eq.\,(11) contributes to the formation of the vertical structure
at the center (around $a_0= 30.22$\,au). Another interesting
resonance is C2, in which the nodal precessions ($s, s_8$ and
$s_6$) are involved. In fact, as the inclination of Trojan orbit
increases, it is natural to expect that the nodal precession is
becoming more important. We would like to note that there are less
regular indentations corresponding to this resonance in the
dynamical map at $a_0=30.08, 30.36$\,au and $e_0\ge 0.015$. On the
other hand, although the resonances C3, C4 in Eq.\,(11) can be
detected in the dynamical spectra, no apparent details in the
dynamical map can be found related to these resonances.

The influences from the C-type resonances are less important in
slice $i_0=35^\circ$ than in slice $i_0=5^\circ$. Relatively, the
effects from the S-type resonances are remarkable. Beside the
$\nu_8, \nu_{18}$ and Kozai resonances that locate outside the
survival region as show in Fig.\,6, several other S-type
resonances as below are determined and illustrated.
\begin{equation}
\begin{array}{ll}
{\rm S1}: & g+s-g_8-s_8=0, \\
{\rm S2}: & g-2s+g_8-s_8+s_5=0, \\
{\rm S3}: & g+2s-g_6+s_6-3s_5=0, \\
{\rm S4}: & g-2s+g_8+s_8+g_7-g_5-s_5=0.
\end{array}
\end{equation}
Clearly the S2 resonance combined with S3 defines the boundary of
the survival region, and the resonance S4 establishes the edge of
the stable region up to $e_0=0.12$. Again, on the top part of the
dynamical map, the C-type and S-type resonances gather, resulting
in overlaps among them and leading to chaotic motion.

\subsection{For $i_0=55^\circ$}
Setting $\delta= a_0-30.223$, the formulae of the proper
frequencies for the slice $i_0=55^\circ$ are:
\[
 f_\sigma [10^{-4}\,2\pi{\rm /yr}]=0.88663-1.1998\delta^2-2.0233e^2
\]
\begin{equation}
\\-7.4512\delta^4 +353.60e^4-106.00\delta^2e^2,
\end{equation}
\[
g [10^{-7}\,2\pi{\rm /yr}]=2.7724-46.683\delta^2-71.157e^2
\]
\begin{equation}
 \\-1588.8\delta^4 +666.87e^4-3776.5\delta^2e^2,
\end{equation}
\[
 s [10^{-7}\,2\pi{\rm /yr}]=1.5643-46.165\delta^2-52.989e^2
\]
\begin{equation}
\\+2168.4\delta^4+5116.5e^4+18800\delta^2e^2.
\end{equation}

\begin{figure}
\label{fig07}
 \vspace{7.5cm}
 \includegraphics{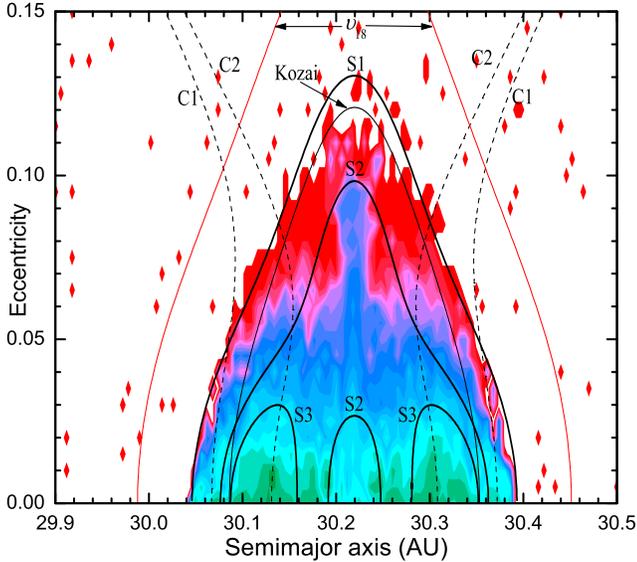}
 \caption{The same as Fig.\,5 but for $i_0=55^\circ$. Except the resonances
 labeled by C1, C2 (dashed) and S1-S3 (solid. See text for the meanings), the locations of
 $\nu_{18}$ resonances (red lines) and Kozai resonance (thin solid curve in black)
 have also been plotted.}
\end{figure}

The involved important resonances are determined and illustrated
in Fig.\,7, where the following C-type and S-type resonances are
plotted:
\begin{equation}
\begin{array}{ll}
{\rm C1}: & 2f_\sigma -f_{2:1} -2g+ 3g_6=0, \\
{\rm C2}: & 2f_\sigma -f_{2:1} -g -s + 3s_6=0,\\
{\rm S1}: & 2s-2g_5-2s_5+g_6-s_6+g_7+s_7=0, \\
{\rm S2}: & s-2g_7+2s_7-s_5=0, \\
{\rm S3}: & s-g_8-s_8-g_7+g_5+s_5=0.
\end{array}
\end{equation}
Although the most part of the C1 lines in Fig.\,7 are out of the
survival region, we see the fine structures (narrow unstable gaps)
around $a_0=30.06, 30.36$\,au and $e_0>0.002$ are clearly related
to this resonance. Inversely, the stability of orbits benefits
from the C2 resonance, as we see the concentration of stable
orbits around the C2 lines, especially in the regions $a_0=30.13,
30.31$\,au and $e_0<0.002$.

Comparatively, the dynamical effects from S-type resonances are
more prominent. As shown in Fig.\,7, the S1 resonance sharply
defines the out edge of the survival region; the lines of S2
coincide with the boundary separating the chaotic orbits from the
relatively regular ones; and the S3 curves surround the most
stable orbits in this slice.

Considering the high inclination of this slice at $i_0=55^\circ$, it
is not a surprise to see that the nodal precession $s$ is involved
more deeply in these important resonances compared to the low
inclination cases (e.g., resonances listed in Eqs.\,(6), (7), (11)
\& (12)).

Meanwhile, the $\nu_{18}$ resonance (red lines) and Kozai resonance
(thin black line) are calculated and shown in Fig.\,7. The
$\nu_{18}$ is outside the survival region, that is, it does not play
an important role in the long-term stability of orbits. If an orbit
was trapped in Kozai resonance, its inclination $i$ and eccentricity
$e$ may experience joint variations: $i$ decreases as $e$ increases,
or vice versa \cite{koz62}. This variation may lead to a highly
eccentric orbit and result in orbit escaping. Therefore, the
boundary between the surviving and escaping orbits is determined
cooperatively by the S1 and Kozai resonances.

\subsection{A summary}
At different inclinations, different resonances participate in
featuring the dynamical maps. Basically, the secondary resonances
(C-type) related to the libration of the 1:1 MMR between Neptune
and its Trojans and to the quasi 2:1 MMR between Neptune and
Uranus, as well as the secular resonances (S-type) related to the
nodal and/or apsidal processions, are the most important dynamical
mechanisms that form the fine details in the dynamical maps and
determine the complicated orbital behavior of NTs. Compared to the
cases in low inclination, those highly inclined orbits are more
deeply affected by the nodal-type resonances. In all cases,
different resonances with different orders make a resonance net in
the parameter plane of initial conditions. Thus an orbit may
diffuse slowly along the net and migrate on the parameter space
from one configuration to another. Of particular interest is the
slow diffusion from low inclination orbits to high inclination
orbits, which can be used to explain the origins of the observed
NTs with high-inclinations, i.e. 2005 TN53 and 2007 VL305 as
listed in Table 1.

We only reported in previous subsections some resonances detected by
the dynamical spectrum method. Surely, many more resonances,
particularly those with high orders, have not been revealed in our
current paper, and they may also play important roles in
characterizing the dynamical behaviours of Trojans' orbits, which
are our studying topic in future.

\section{Orbits of Observed Neptune Trojans}
Up to now, six NTs have been observed, and their orbital
stabilities have been analyzed individually, e.g. in
\cite{mar03a,bra04,lij07}. Since we have presented global view of
the dynamics of NTs, we can locate the observed NTs on the
corresponding dynamical maps and draw conclusions directly from
the comparison.

We have adopted the configuration of the outer solar system at
epoch of August 1, 1993 (JD2449200.5) as the initial conditions to
simulate the orbital evolutions of four planets and artificial
Trojans, the dynamical maps (Fig.\,3 in this paper and also
Figs.\,2\,\&\,3 in Paper I) are based on the instantaneous orbital
elements at that moment. But the orbits of observed NTs are given
in recent epoch, e.g. the ones listed in Table 1 obtained from the
AstDyS website\footnote{http://http://hamilton.dm.unipi.it/astdys}
are given at epoch JD2455000.5. To locate their orbits on the
dynamical maps, we need to transfer their orbital elements at this
moment to the epoch when the planets' orbits are adopted. The
transferred orbits are calculated and listed in Table 1.

\begin{table*}
\caption{Orbits of observed Neptune's Trojans at epochs JD=2455000.5
(June 14, 2009, six columns in the left half) and JD=2449200.5
(August 1, 1993, six columns in the right half) in the reference
frame of J2000.0. The mean anomaly $M$, perihelion argument
$\omega$, ascending node $\Omega$ and inclination $i$ are in
degrees. All the angle values have been rounded to one decimal place
for simplicity. The last column $\Delta\sigma$ gives the
approximations of the libration amplitude (in degrees) of $\sigma$,
derived from Fig.\,2 in Paper I. }
 \center{\begin{tabular}{|l|r|r|r|r|c|r|c|r|r|r|r|c|r|c|}
 \hline
 Designation & $M$ & $\omega$ & $\Omega$ & $i$ & $e$ & $a$\,(AU) & $\rightarrow$ & $M$ & $\omega$ & $\Omega$ & $i$ & $e$ & $a$\,(AU) & $\Delta\sigma$ \\
 \hline
 2001 QR322 &  57.31 & 162.5 & 151.6 &  1.3 & 0.0316 & 30.336 & $\rightarrow$ &  30.06 & 155.3 & 151.6 &  1.3 & 0.0290 & 30.208 & 77 \\
 2004 UP10  & 343.40 & 357.5 & 34.8  &  1.4 & 0.0295 & 30.250 & $\rightarrow$ & 303.65 &   2.8 &  34.7 &  1.4 & 0.0259 & 30.121 & 48 \\
 2005 TN53  & 289.14 &  84.8 &  9.3  & 25.0 & 0.0652 & 30.213 & $\rightarrow$ & 250.71 &  88.7 &   9.3 & 25.1 & 0.0652 & 30.099 & 33 \\
 2005 TO74  & 270.02 & 301.7 & 169.4 &  5.2 & 0.0501 & 30.222 & $\rightarrow$ & 230.28 & 306.9 & 169.4 &  5.2 & 0.0517 & 30.095 & 37 \\
 2006 RJ103 & 242.91 &  24.2 & 120.8 &  8.2 & 0.0276 & 30.112 & $\rightarrow$ & 202.77 &  29.5 & 120.9 &  8.2 & 0.0318 & 29.990 & 22 \\
 2007 VL305 & 354.05 & 215.3 & 188.6 & 28.1 & 0.0648 & 30.081 & $\rightarrow$ & 317.23 & 217.5 & 188.6 & 28.2 & 0.0620 & 29.982 & 20 \\
 \hline
 \end{tabular}}
\end{table*}

Because all the six objects are on nearly circular orbits with
eccentricities smaller than 0.07, we can approximately locate
these orbits on the $(a_0,i_0)-$\,plane in Paper I
(Figs.\,2\,\&\,3) which are made for artificial Trojans with
initial eccentricity $e_0\sim 0$. Since they are all around the
leading Lagrange point ($L_4$), we should put them on Fig.\,2 in
Paper I (for $L_4$) to compare. A comparison tell clearly that
five of them are inside the most stable regions (Regions A and B)
and the only exception is 2001 QR322 (hereafter QR322 for short),
whose position in the dynamical map is near the low-inclination
end of the ``arc structure''. We have shown in Paper I that this
arc structure is related to a resonance as $f_{2:1} - 2f_\sigma =
g_6$, i.e., C1 in Eq.\,(6) of this paper.

A more accurate comparison can be performed by using the dynamical
maps at different inclinations in this paper (e.g. Fig.\,3).
Recalling that all dynamical maps in this paper are for orbits
around the trailing Lagrange point ($L_5$), we cannot put the
observed orbits directly on the corresponding maps. Nevertheless,
since the absolute symmetry between $L_4$ and $L_5$ have been
proven \cite{nes02a,mar03a,zho09a}, it is possible to find the
symmetrical counterpart around $L_5$ point for each orbit. To do
so, we read from Fig.\,2 in Paper I the libration amplitude
$\Delta\sigma$ for each observed NT, which have been listed in the
last column of Table 1. Then, remembering the inclination and
neglecting the semimajor axis, the corresponding counterpart
around $L_5$ is the one librating with the same $\Delta\sigma$
value on the $(a_0,e_0)-$\,plane with the same inclination. For
example, the object QR322 has a small inclination ($1.3^\circ$),
and approximately we regard it as on the slice of $i_0=0^\circ$,
i.e. the first panel in Fig.\,3. We check the libration
($\Delta\sigma=77^\circ$) and find that the best counterpart of
QR322 is $(a_0,e_0)\approx(30.38\,{\rm au}, 0.290)$. It is on the
stable segment to the right of the unstable gap at
$a_0\sim30.375$\,au. Similarly, we can locate the symmetrical
correspondence of 2005 TO74 (hereafter TO74) in Fig.\,5 for
$i_0=5^\circ$ at $(a_0,e_0)\approx(30.26\,{\rm au}, 0.052)$. It is
on the stable side of the boundary between stable and chaotic
motion. Moreover, it's in the very vicinity of the S3 resonance
curve. We may conclude that the orbits of both QR322 and TO74 are
stable and these objects, however, are probably influenced by
resonances like C1 in Eq.\,(6) and S3 in Eq.\,(7), respectively.
The rest four objects listed in Table 1, again this time, are well
inside the most stable regions. Thus we neglect here the
comparisons in detail.

\begin{figure}
\label{fig08}
 \vspace{7.5cm}
 \includegraphics{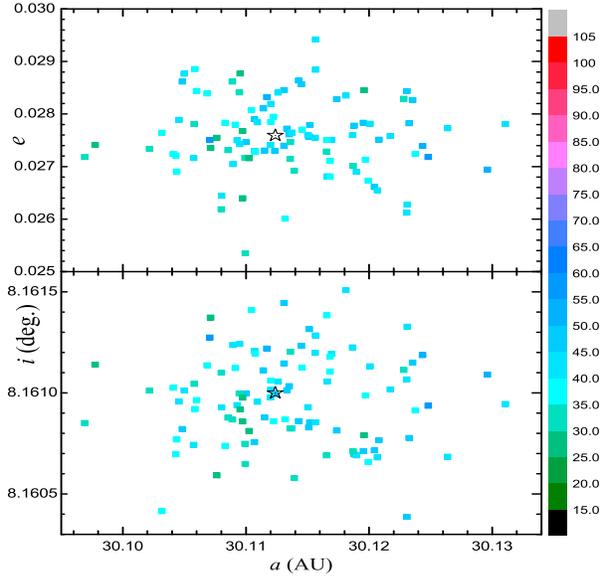}
 \caption{The orbital stabilities of 100 clones of 2006 RJ103. The color
 indicates the spectral number that is used as the stability indicator.
 The black stars designate the nominal orbital elements on the $(a,e)$
 and $(a,i)$ plane (upper and lower panel respectively).}
\end{figure}

However, the observed orbits are not exactly on the slices we have
calculated. To confirm our conclusion and to get more accurate
estimation of the orbital stabilities, we integrate one hundred
clones of each nominal orbit (note the numbers listed in Table 1
have been rounded for simplicity) and check their stabilities using
the frequency analysis method adopted in our papers. Considering the
uncertainties from observation and orbital determination, the clone
orbits are generated using the covariance matrix given in the AstDyS
webpage. We thank Dr. Antonio Giorgilli for supplying us the
necessary computing codes for generating the proper initial
conditions.

The results confirm the orbital stabilities of 2004 UP10, 2005 TN53,
2006 RJ103 and 2007 VL305. They are deeply inside the stable region,
as proven by the fact that nearly all the clones of them have
SN$<50$. As example, we present in Fig.\,8 the SN of clones of 2006
RJ103. Out of the 100 orbits, only 4 have SN$>50$ with the largest
SN$=57$.

As mentioned above, TO74 is close to the edge of stable region.
Considerable fraction ($\sim 50\%$) of the clones has relatively
large SN$\sim 70$. And there is a trend of declining SN as the
eccentricity of clones decreases. Thus we may expect that further
observations will constrain the eccentricity to a smaller value.

\begin{figure}
\label{fig09}
 \vspace{7.5cm}
 \includegraphics{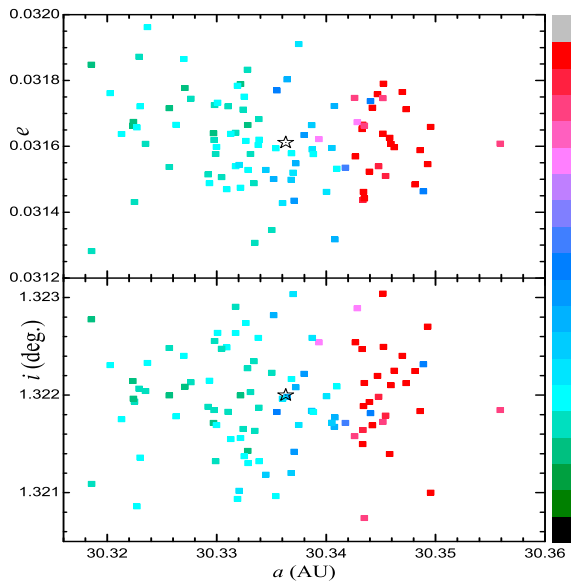}
 \caption{The same as Fig.\,8, but for object 2001 QR322.}
\end{figure}

The most interesting result is from the case for QR322, as shown
in Fig.\,9. Because QR322 is in a narrow stable ``stripe-like''
area separated from the main stable region in the dynamical map,
some clones of this object are protected by the C-type resonance
$f_{2:1} - 2f_\sigma = g_6$, while some other clones nearby may
suffer the destroying effects from resonance like $2f_\sigma -
f_{2:1} - g + g_8 + g_6=0$ and its analogues. In Fig.\,9, we see
the stability of clones depends sensitively on the semimajor axis,
which can be explained by the sensitive dependence of the
libration frequency $f_\sigma$ on semimajor axis. The eccentricity
and inclination within the observing error range, on the other
hand, do not influence the orbital stability considerably. Similar
results can also be found in \cite{hor10b}, where the lifetime of
clones has a sharp edge on semimajor axis (Figs.\,3\,\&\,4
therein). In this sense, the QR322 is not a typical (thus not a
good) example of NTs. If using it as a seed to generate a cloud of
artificial asteroids around to represent the real NTs
\cite{hor10a}, additional attention should be paid to make the
conclusion drawn from the orbital simulations of the cloud
consistent and convictive.

In addition, we have integrated the six nominal orbits up to the
Solar system age (4.5\,Gyr) and found that all of them survive on
the Trojan-like orbits. According to our investigations above, it
is an expectable outcome for five objects. But for QR322, we would
like to say, it is more or less by luck. Not like others, only a
very slight deviation from the nominal orbit of QR322 may lead to
an unstable orbit.

\begin{figure}
\label{fig10}
 \vspace{10.5 cm}
 \includegraphics{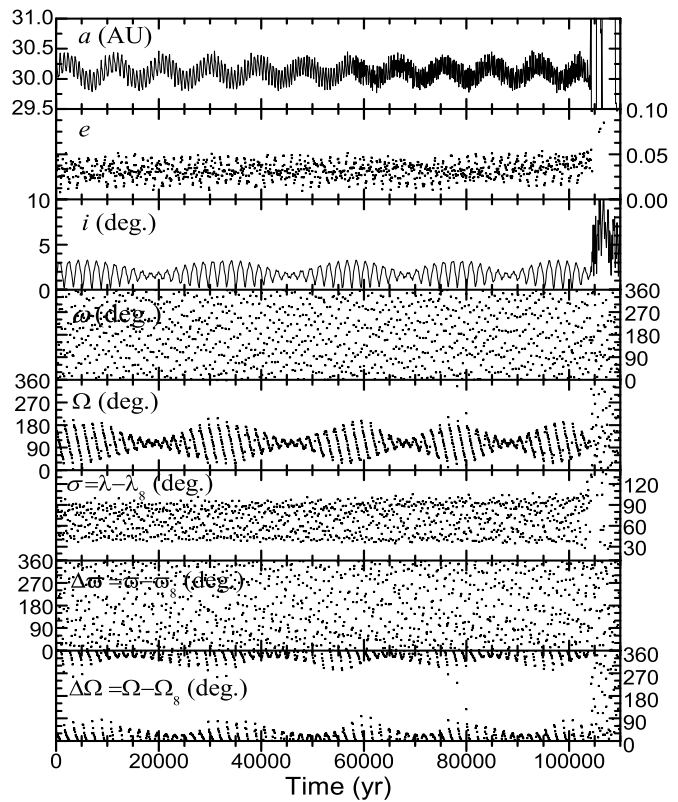}
 \caption{The orbital evolution of a clone of 2001 QR322. From top
 to bottom, we illustrate in 8 panels the temporal evolution of $a, e, i,
 \omega$ (critical angle of Kozai resonance),
 $\Omega, \sigma = \lambda-\lambda_8$ (resonant angle of the 1:1 mean
 motion resonance), $\Delta\varpi= \varpi-\varpi_8$ (critical angle
 of $\nu_8$ resonance) and $\Delta\Omega = \Omega - \Omega_8$ (
 critical angle of $\nu_{18}$ resonance).}
\end{figure}

In Fig.\,10, we show such an example. The initial orbit is exactly
the literal one listed in Table 1, in which the numbers have been
rounded for simplicity, e.g. the inclination $1.3^\circ$ is
rounded from the nominal value $1.3220^\circ$. At the beginning
$1\times 10^5$\,yrs, the orbit has a quite regular behaviour. The
eccentricity is small, the inclination is constrained below
$4^circ$, and the resonant angle $\sigma$ varies with an amplitude
smaller than $70^\circ$. In most of the time, it is trapped in the
$\nu_{18}$ resonance, as $\Delta\Omega$ librates around $0^\circ$
with an amplitude smaller than $180^\circ$. Then from $1.05\times
10^5$\,yrs, the eccentricity increases and the orbit loses
stability, escaping from the Trojan-like orbit.

A close observation at Fig.\,10 may reveal that as early as $4\times
10^4$\,yrs, some chaotic features can be found in the semimajor
axis' behaviour, and the eccentricity begins to show a slow
increasing trend from $8\times 10^4$\,yrs. We think it is due to the
fact that this orbit is in the resonance $2f_\sigma - f_{2:1} - g +
g_8 + g_6=0$. The libration $f_\sigma$ and apsidal precession $g$
are involved in this resonance, therefore the chaos was introduced
to the behaviour of $a$ and $e$. Finally, the resonance results in
totally chaotic orbit.

\section{Conclusions}
As a continuation of our research in a previous paper on the
dynamics of NTs, we present in this paper a detailed investigation
on the stability of orbits on the initial plane of $(a_0,e_0)$. By
constructing dynamical maps on slices with different inclinations
$i_0$, we obtain a global view of the stability of NTs in the
whole orbital parameter space.

In the three most stable regions in inclination, we found in the
representative dynamical maps the extension of stable motion in
eccentricity is 0.10 for low-inclination orbits with
$i_0<12^\circ$, 0.12 for medium inclination $22^\circ < i_0 <
36^\circ$, and 0.04 for high inclination between $51^\circ$ and
$59^\circ$. The feature of dynamical map was enriched by fine
structures in it, indicating the diversity of orbital behaviour of
NTs.

Using the dynamical spectrum method based on the frequency
analysis, we figured out the mechanisms creating the fine
structures in the dynamical maps. We found two types of resonance
may involve deeply in the dynamics of NT. One is the secondary
resonance (C-type) concerning mainly two frequencies: the
libration frequency of the resonant angle $\sigma= \lambda-
\lambda_8$ and the frequency of the quasi 2:1 MMR between Neptune
and Uranus $2\lambda_8 - \lambda_7$. The other is the secular
resonance in general sense (S-type), characterized by the
commeasurability between different combinations of secular
frequencies, such as the frequencies of apsidal and/or nodal
precessions of NT and/or planets. Among the well-known secular
resonances, the $\nu_8$ resonance is very effective in driving a
Trojan to highly eccentric orbit thus make a deep unstable gap in
inclination around $44^\circ$. On the contrary, the $\nu_{18}$
resonance, found nearly everywhere at low inclination, is so weak
that it has hardly any influence on the dynamics of NT.

Our study is not just theoretical, since we can place the observed
NTs on the dynamical map and check whether they are trapped in or
close to some identified resonances. In this way, the dynamical
features of the observed objects can be predicted. We found 2004
UP10, 2005 TN53, 2006 RJ103 and 2007 VL305 are deeply inside the
region of the most regular motion, far away from dominant
destructive secular resonances. Therefore they must be stable. The
orbits of 2001 QR322 and 2005 TO74 may survive to the Solar system
age, but they are probably influenced by specific C-type and
S-type resonances, respectively. And future observations may
constrain their orbits to stable region a little deeper. To
confirm the above conclusions, hundreds of orbital clones of the
six observed objects were generated within the observational error
bars, their orbits were integrated numerically, and their
stabilities were carefully inspected.

Surely, the resonances listed in Eqs.\,(6), (7), (11), (12), (16)
are not all of the possible resonances involved in the orbital
evolution. Especially, we may miss some important resonances with
higher order. However, even only these resonances listed in our
paper, have built a ``resonant net'' that connect different parts
of the whole orbital parameter space. Correspondingly, there must
be such a net in the phase space. With the help of this net, an
orbit may diffuse in the phase space, resulting in significant
change in the orbital behaviour. Long-term diffusion may be
expectable along the net. The most attractive diffusion, probably
is the slow transferring of an orbit from low inclination to high
inclination. Although in our preliminary tests, we did not find
such inclination-increasing phenomenon, we would argue that this
is a very important issue. The importance arises not only from the
interesting dynamics itself, but also from the fact that this
problem is closely related to the origins of those highly inclined
NTs. Again, their origins are closely related to the evolution of
our planetary system in the early stage \cite{kor04,nes09}.

Another possible hint about the early evolution of the Solar
system is buried in the C-type resonances. On one hand, C-type
resonances are quite powerful in featuring the dynamics of NTs. On
the other hand, these resonances are sensitively affected by the
frequency $f_{2:1}$. If the planetary orbital configuration
changed, even only very little, $f_{2:1}$ would change and this
would lead to significant varying in the position and strength of
the C-type resonances. Thus a careful discussion on the influences
of the C-type resonances, under current planetary configuration
and under other possible configuration, is critical. Similar
analysis have been performed in the case of the formation of the
Hecuba Gap in the main asteroid belt, where the Great Inequality
(quasi 5:2 MMR between Jupiter and Saturn) plays the role of
$f_{2:1}$ here \cite{hen97}. A closer analogue is about Jupiter
Trojans \cite{rob09}. For NTs, some work have been done
\cite{nes09,lyk10}, and our contribution is also undergoing.

\section*{Acknowledgements}
This work was supported by the Natural Science Foundation of China
(No. 10403004, 10833001, 10803003), the National Basic Research
Program of China (2007CB814800). L.Zhou thanks University of Vienna
for the financial support during his stay in Austria.

\end{document}